\documentclass[ALICE,manyauthors]{cernphprep}
\usepackage[comma,square,numbers,sort&compress]{natbib}
\usepackage{hyperref}
\usepackage{graphicx}
\usepackage{dcolumn}
\usepackage{bm}
\usepackage{amssymb}
\usepackage{amsfonts}
\usepackage{graphics}
\usepackage{grffile}
\usepackage{epsfig}
\usepackage{mathptmx}
\usepackage{units}
\usepackage[usenames]{color}
\usepackage[normalem]{ulem}
\usepackage{color}
\usepackage[utf8]{inputenc}
\usepackage[T1]{fontenc}
\usepackage{orcidlink}
\usepackage{subfigure}
\usepackage{lineno}
\usepackage{tabularx,booktabs}

\usepackage{xspace}
\usepackage{color}
\usepackage{xcolor}
\usepackage{rotating}
\definecolor{light-gray}{gray}{0.8}

\newcommand{\cmnt}[1]{}

\newcommand{\pt}{\ensuremath{p_{\rm T}}\xspace}
\newcommand{\mt}{\ensuremath{m_{\rm T}}\xspace}

\newcommand{\sqrtsNN}{\ensuremath{\sqrt{s_{\rm NN}}}\xspace}

\newcommand{\dndy}{\ensuremath{{\rm d}N/{\rm d}y}\xspace}

\newcommand{\dnchdetacubic}{\ensuremath{\average{{\rm d}N_{\rm ch}/{\rm d}\eta}^{1/3}}\xspace}

\newcommand{\etaless}[1]{\ensuremath{\left|\eta\right| < #1}\xspace}
\newcommand{\yless}[1]{\ensuremath{\left|y\right| < #1}\xspace}

\newcommand{\GeVc}{GeV/$c$\xspace}

\newcommand{\fmc}{\rm fm/$c$\xspace}

\newcommand{\PbPb}{Pb--Pb\xspace}

\newcommand{\dAu}{d--Au\xspace}
\newcommand{\AuAu}{Au--Au\xspace}

\newcommand{\average}[1]{\ensuremath{\langle #1 \rangle}\xspace}
\newcommand{\dedx}{\ensuremath{{\rm d}E/{\rm d}x}\xspace}

\newcommand{\pKminus}{\ensuremath{{\rm K}^{-}}\xspace}

\newcommand{\pProtonNS}{\ensuremath{\mathrm{p}}}

\newcommand{\pLambda}{\ensuremath{\Lambda}\xspace}

\newcommand{\Lambdastar}{\ensuremath{\Lambda(1520)}\xspace}

\newcommand{\Kstar}{\ensuremath{{\rm K}^{\ast}(892)^{0}}\xspace}

\newcommand{\Kstarpm}{\ensuremath{{\rm K}^{\ast}(892)^{\pm}}\xspace}
\newcommand{\rhozero}{\ensuremath{\mathrm{\rho}(770)^{0}}\xspace}

\begin{document}

\begin{titlepage}

\PHyear{2026}       
\PHnumber{302}      
\PHdate{11 November}  

\title{ Observation of the \Lambdastar suppression in \PbPb collisions at $\mathbf{\sqrt{s_{\rm NN}}}$~=~5.02 TeV}

\ShortTitle{ $\Lambda(1520)$ in \PbPb collisions at $\sqrt{{s}_{\textbf{NN}}}$ = 5.02 TeV}   

\Collaboration{ALICE Collaboration\thanks{See Appendix~\ref{app:collab} for the list of collaboration members}}
\ShortAuthor{ALICE Collaboration} 

  \begin{abstract}
   A suppression of the baryonic \Lambdastar resonance yield relative to the \pLambda ground state by $0.589 \pm 0.015^\mathrm{stat} \pm 0.086^\mathrm{syst}$ is observed in central relative to peripheral \PbPb collisions, with a significance of $6.8\sigma$.
The \Lambdastar production at midrapidity is measured in six centrality classes in \PbPb collisions for the first time at \sqrtsNN~=~5.02~TeV with ALICE. 
The \Lambdastar is a sensitive probe of late-stage hadronic interactions since its lifetime is comparable to that of the hadronic phase in heavy-ion collisions at the LHC. 
Statistical hadronization model calculations, predicting yields at chemical freeze-out, overestimate the measured ratio.
The transverse-momentum spectra and the centrality dependence of the suppression are reproduced by the MUSIC hydrodynamic model only when coupled to the SMASH hadronic transport afterburner, which implements rescattering and regeneration in the late hadronic stage. 
The measurement, which represents the first observation of resonance suppression within a single collision system, provides a stringent test of the late-stage dynamics and evidence that hadronic interactions substantially modify reconstructible resonance yield.
  \end{abstract}

\end{titlepage}

\setcounter{page}{2} 


Lattice quantum chromodynamics (QCD) calculations predict that ordinary hadronic matter undergoes a transition into a deconfined phase of quarks and gluons, the quark–gluon plasma (QGP)~\cite{Cabibbo:1975ig}, when the temperature exceeds the pseudo-critical value $T_{\mathrm {pc}}\approx 155$ MeV~\cite{Bazavov:2018mes,Borsanyi:2020fev}. Heavy-ion collisions at ultra-relativistic energies provide the only means to reproduce such extreme conditions in the laboratory, offering unique access to the properties of strongly-interacting matter.
At the LHC, the temperature reached in the collision region significantly surpasses $T_{\mathrm {pc}}$~\cite{ALICE:2022imr,ALICE:2022wpn}. In this regime, the formed QGP exhibits the characteristics of a nearly perfect fluid that evolves in an approximate local equilibrium. As the system expands and cools, the temperature eventually drops below $T_{\mathrm {pc}}$, triggering the hadronization process that turns it into confined matter~\cite{Bazavov:2018mes}. This partonic stage is followed by a hadronic gas phase~\cite{Muller:2006ee} with temperatures in the range of about 110–150 MeV~\cite{ALICE:2019hno,Mazeliauskas:2018irt,Mazeliauskas:2019ifr}.
During this phase, inelastic interactions gradually cease, fixing the final hadron abundances at the chemical freeze-out point, characterized by the temperature $T_{\rm ch}$ and baryochemical potential $\mu_{\rm B}$~\cite{Andronic:2005yp,Andronic:2011yq}.
Afterwards, hadrons and their resonances continue to interact (pseudo)elastically until the system arrives at the kinetic freeze-out point, characterized by the temperature $T_{\rm kin}$.
During its evolution, the medium develops a collective transverse expansion, known as radial flow, which shifts hadron transverse-momentum distributions towards higher \pt, with a stronger effect for heavier particles.

Two- and three-pion femtoscopic measurements in central Pb--Pb collisions at $\sqrt{s_{\mathrm{NN}}}=2.76$ TeV indicate a characteristic decoupling proper time of approximately 10–11 \fmc within a longitudinal-expansion model~\cite{ALICE:2011dyt,ALICE:2022wpn,ALICE:2013uhj}. This time scale encompasses the evolution preceding pion decoupling and should be distinguished from the duration of the hadronic phase alone.
The characteristic short lifetime ($\tau \approx$ few \fmc) of hadronic resonances makes them excellent probes of the hadronic phase created in heavy-ion collisions. 
Due to their short life, resonances may decay before kinetic freeze-out, and the products (daughter particles) may interact with the neighbouring particles that populate the hadron gas. Re-scattering can destroy the original correlations of daughters and decrease the resonance yield observed experimentally.
At the same time, a close correlation of nearby hadrons can lead to the creation or regeneration of a new resonance, which increases the observed yield.
Both of these phenomena are affected by the duration of the hadronic phase and the particle density, the hadronic interaction cross sections of the decay daughters inside the hadronic medium, and the resonance lifetime. 
Short-lived resonances such as \rhozero~\cite{ALICE:2018qdv}, \Kstarpm~\cite{ALICE:2023ifn}, and \Kstar~\cite{ALICE:2019xyr, ALICE:2021ptz} ($\tau \approx$ 1.34, 4, 4.16 \fmc, respectively) are found to exhibit a suppressed yield relative to long-lived particles in central \PbPb collisions compared to peripheral collisions following a multiplicity dependent suppression, which is attributed to the dominance of the rescattering of decay products over the regeneration of resonances inside the hadronic medium created in central collisions~\cite{ALICE:2019xyr}.
Conversely, the yield of long-lived $\phi(1020)$ ($\tau \approx 46.4$ \fmc~\cite{ParticleDataGroup:2024cfk}) does not show any suppression across centralities, which indicates that it decays outside of the fireball, where the decay products are no longer subject to rescattering in the medium.
The study of the \Lambdastar state is extremely valuable owing to its intermediate lifetime of $\tau =(~12.55 \pm ~0.21)$~\fmc~\cite{ParticleDataGroup:2024cfk}. Results from the ALICE collaboration in \PbPb collisions at \mbox{\sqrtsNN = 2.76 TeV}~\cite{ALICE:2018ewo} provided the first indication of a suppressed yield of short-lived \Lambdastar in central collisions with respect to peripheral collisions with a 3.1$\sigma$ significance. 

This Letter presents the first observation of suppression of the $\Lambdastar/\Lambda$ yield ratio in \PbPb collisions at \sqrtsNN = 5.02 TeV with the ALICE detector at the LHC. Results are compared with predictions from statistical hadronization and hydrodynamic models, including calculations with and without hadronic transport, to investigate the role of late-stage hadronic interactions in $\Lambdastar$ production.

The complete ALICE experimental setup and performance are described in detail in Refs.~\cite{ALICE:2008ngc,Abelev:2014ffa,ALICE:2022wpn}. Only the detectors relevant for this analysis are briefly summarized here. The V0A and V0C scintillator hodoscopes~\cite{ALICE:2013axi}, covering $2.8 < \eta < 5.1$ and $-3.7 < \eta < -1.7$, respectively, measure particle arrival times with 1 ns resolution and are used for triggering, beam–background suppression, and centrality determination.
The Inner Tracking System (ITS)~\cite{ALICE:2010tia}, a six-layer high-resolution silicon tracker, together with the Time-Projection Chamber (TPC)~\cite{Alme:2010ke}, a large cylindrical drift detector, provides vertex reconstruction and charged-particle tracking in ALICE’s central barrel within the 0.5 T solenoidal magnetic field of the L3 magnet. The Time-Of-Flight system (TOF)~\cite{ALICE:2000xcm,ALICE:2002imy}, a large array of multigap resistive plate chambers placed at a radius of 370--399 cm from the beamline, contributes to charged-particle identification by measuring time-of-flight with $\sim$ 56 ps resolution in LHC Run 2~\cite{Carnesecchi:2018oss}. Along with the ITS and TPC, TOF covers $\vert \eta\vert < 0.9$ and full azimuth. Particle identification (PID) is performed using the TPC's specific ionization energy loss (\dedx), with $\sim$ 5\% resolution for tracks with 159 clusters and $\sim$6.5\% on average for all tracks in 0–5\% central \PbPb collisions, enabling K/p separation at 3$\sigma$ up to $\sim$1 \GeVc. At higher momenta, TOF measurements extend the 3$\sigma$ K/p discrimination up to $p \approx 5$ \GeVc~\cite{Jacazio:2018slq}.

The analysis uses the complete 5.02 TeV \PbPb collision dataset from LHC Run 2, comprising about $4.30 \times 10^{8}$ events: minimum-bias (MB) triggered events in 2015 ($\mathcal{L} = 56~ \mu\mathrm{b}^{-1}$) and 2018, with central (0–10\%, $\mathcal{L} = 105~\mu\mathrm{b}^{-1}$), semi-central (30–50\%, $\mathcal{L} = 51~\mu\mathrm{b}^{-1}$), and MB ($\mathcal{L} = 22~\mu\mathrm{b}^{-1}$) triggers.
The MB trigger requires simultaneous hits in V0A and V0C detectors. Those signal amplitudes scale with event multiplicity and provide the collision centrality estimate~\cite{ALICE:2015juo}.
The MB trigger selects hadronic collisions in 0–90\% centrality, while central and semi-central triggers select 0–10\% and 30–50\% central events, respectively.
Beam-induced and electromagnetic backgrounds are removed offline~\cite{ALICE:2015juo,ALICE-PUBLIC-2015-008,Abelev:2014ffa}, and only events with reconstructed vertex position along the beam line within $V_z = \pm 10$ cm of the nominal interaction point are retained. After these selections,  about $3.81 \times 10^{8}$ events are analyzed.

The production-yield measurement of \Lambdastar is performed via invariant-mass reconstruction of the hadronic decay $\Lambdastar \rightarrow \pProtonNS \pKminus$ (B.R. = 22.5 $\pm$ 0.5\%), including its charge conjugate (cc.) state~\cite{ParticleDataGroup:2024cfk}, to enhance the statistical significance; it is referred to as \Lambdastar throughout. The candidate kaons and protons are selected from ITS and TPC reconstructed tracks with \pt $>$ 0.15 \GeVc and pseudorapidity \etaless{0.8}, ensuring uniform detector acceptance.
A tight constraint on the impact parameter to the primary vertex is applied to reduce the amount of secondary particles produced by weak decays and interactions of primary particles with detector material.
Additional track-quality selection criteria follow the previous prescriptions~\cite{ALICE:2021ptz}. Kaons and protons are identified by combining the specific energy loss (\dedx) in the TPC and the $\beta$ value derived from the TOF-measured time of flight. The discriminating variable $N_{\sigma}^{i}$ is defined as the difference between the measured signal in detector $i =$ (TPC or TOF) and the expected signal for a given particle hypothesis, divided by the detector resolution ($\sigma_{i}$).
TPC-only information is used to identify kaons up to $p =$ 0.45 \GeVc and protons up to $p =$ 0.8 \GeVc. The $N_{\sigma}^{\rm TPC}$  selection varies as 4.0, 3.0, and 2.0 for kaons at $p =$ 0.2, 0.4, and 0.45 \GeVc, and for protons at $p =$ 0.25, 0.7, and 0.8 \GeVc.
In the range $0.45< p < 0.6$ \GeVc for kaons and $0.8 < p < 1.1$ \GeVc for protons, PID requires $N_{\sigma}^{\rm TOF} \leq 3.0$ and $N_{\sigma}^{\rm TPC}\leq 5.0$, if TOF information is available, otherwise $N_{\sigma}^{\rm TPC}\leq 2.0$ is applied.
For higher momenta, candidates are selected only if both $N_{\sigma}^{\rm TOF} \leq 3.0$ and  $N_{\sigma}^{\rm TPC}\leq 5.0$ are satisfied for the kaon and proton hypotheses. 

The \Lambdastar signal reconstruction, background estimation, invariant-mass fit, and raw yield extraction follow the same procedure as the previous \sqrtsNN = 2.76 TeV analysis~\cite{ALICE:2018ewo}. Raw yields are extracted at midrapidity ($\vert y\vert < 0.5$) in different \pt bins across six centrality intervals 0--10\%, 10--30\%, 30--40\%, 40--50\%, 50--70\%, and 70--90\%.  
Yields are corrected for the pK decay branching ratio, detector acceptance, reconstruction, and PID efficiency, evaluated via Monte Carlo simulations. The events are simulated with the HIJING~\cite{Wang:1991hta} event generator, including injected \Lambdastar and anti-particle states, whereas particle transport through the detector is modeled using GEANT4~\cite{GEANT4:2002zbu,Allison:2006ve}.

\begin{table}
  \caption{Relative systematic uncertainties (\%) of the \pt-differential \Lambdastar yields. 
Quoted ranges correspond to the first and last measured \pt intervals. 
Sources correlated across all centralities are marked with (*).}
 \label{tab:sys_summary}
  \centering
  \begin{tabular*}{\textwidth}{@{\extracolsep{\fill}}lcccccc}
    \toprule
    & 0--10\% & 10--30\% & 30--40\% & 40--50\% & 50--70\% & 70--90\% \\
    \midrule
    Signal description (*) & 4.2 & 4.4 & 4.1 & 3.9 & 3.9 & 4.1 \\
    Background description & 10.2 & 8.2 & 4.6 & 7.3 & 3.8 & 3.1  \\
     Tracking (*) & 0.5 - 3.5 & 0.2 - 3.0 & 0.2 - 3.0  & 0.2 - 3.0 & 0.2 - 3.1 & 0.2 - 3.1 \\
     PID  & 7 & 5.1 & 4.1 & 2.8 & 4.0 & 2.0  \\
     TPC-TOF matching (*) & 1.0 - 7.5 & 0.5 - 7.7 & 0.5 - 7.7 & 0.5 - 7.4 & 0.5 - 7.5 & 0.4 - 8.0 \\
    Hadronic cross section (*) & 2.2 - 2.0 & 2.4 - 1.8 & 2.4 - 1.8 & 2.4 - 1.8 & 2.4 - 1.8 & 2.7 - 2.0  \\
    Material budget (*) & 1.8 - 0.8 & 2.2 - 0.8 & 2.2 - 0.8 & 2.4 - 0.4 & 2.4 - 0.6 & 3.4 - 1.0 \\
    \pt shape correction & 7.0 - 0.0 & 6.0 - 0.0 & 4.0 - 0.0  & 3.0 - 0.0 & 1.0 - 0.0 & 1.0 - 0.0 \\
    \midrule
    Total & 14.8 - 15.3 & 12.6 - 13.4 & 9.0 - 11.3 & 9.8 - 12.0 & 7.6 - 10.7 & 7.0 - 10.4\\
    \midrule
    Uncorrelated & 14.0 - 12.2 & 11.4 - 9.7 & 7.4 - 6.2 & 8.3 - 7.8 & 5.6 - 5.5 & 3.8 - 3.7 \\
    \bottomrule
  \end{tabular*}
\end{table}

Systematic uncertainties of the corrected \Lambdastar yields are summarized in Table~\ref{tab:sys_summary}. 
The total uncertainty is obtained by summing all contributions in quadrature.
These contributions are classified as uncorrelated or correlated across centrality classes. The combined correlated contribution ranges from 5 to 9\%, depending on \pt.

\begin{figure}
\begin{center}
\includegraphics[width=0.5\linewidth]{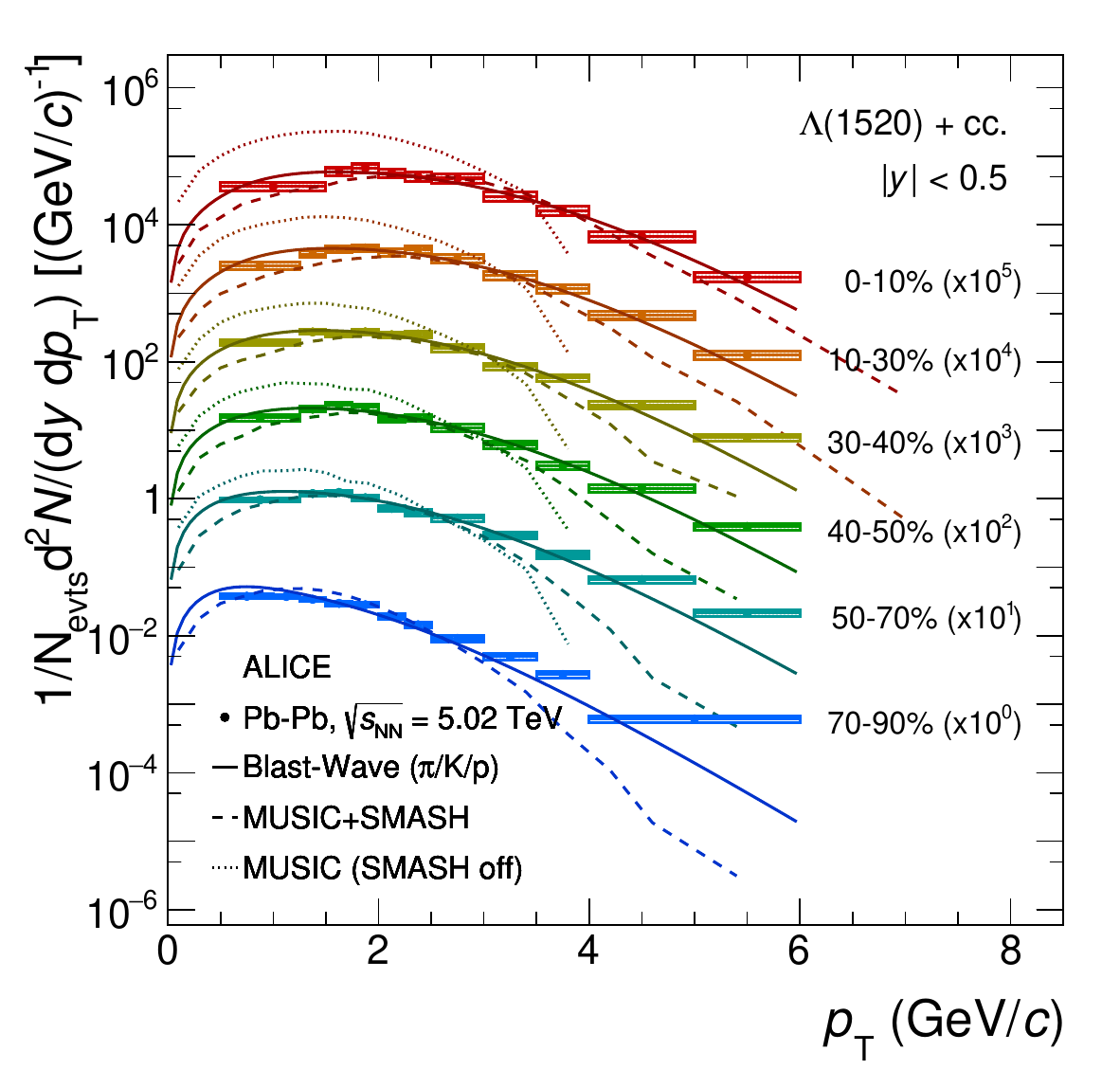}
\caption{The \pt-differential \Lambdastar yields at midrapidity, \yless{0.5}, for six centrality classes. Solid, dashed, and dotted curves show predictions from the Blast-Wave model (normalized to data), MUSIC with and without the SMASH afterburner~\cite{SMASH:2016zqf,Oliinychenko:2021enj}, respectively. Statistical and systematic uncertainties are indicated by bars and boxes, respectively. Shaded boxes indicate uncorrelated systematic uncertainties.
}
\label{fig:spectra}
\end{center}
\end{figure}
%
%
\begin{figure}
\centering
\includegraphics[width=0.5\textwidth]{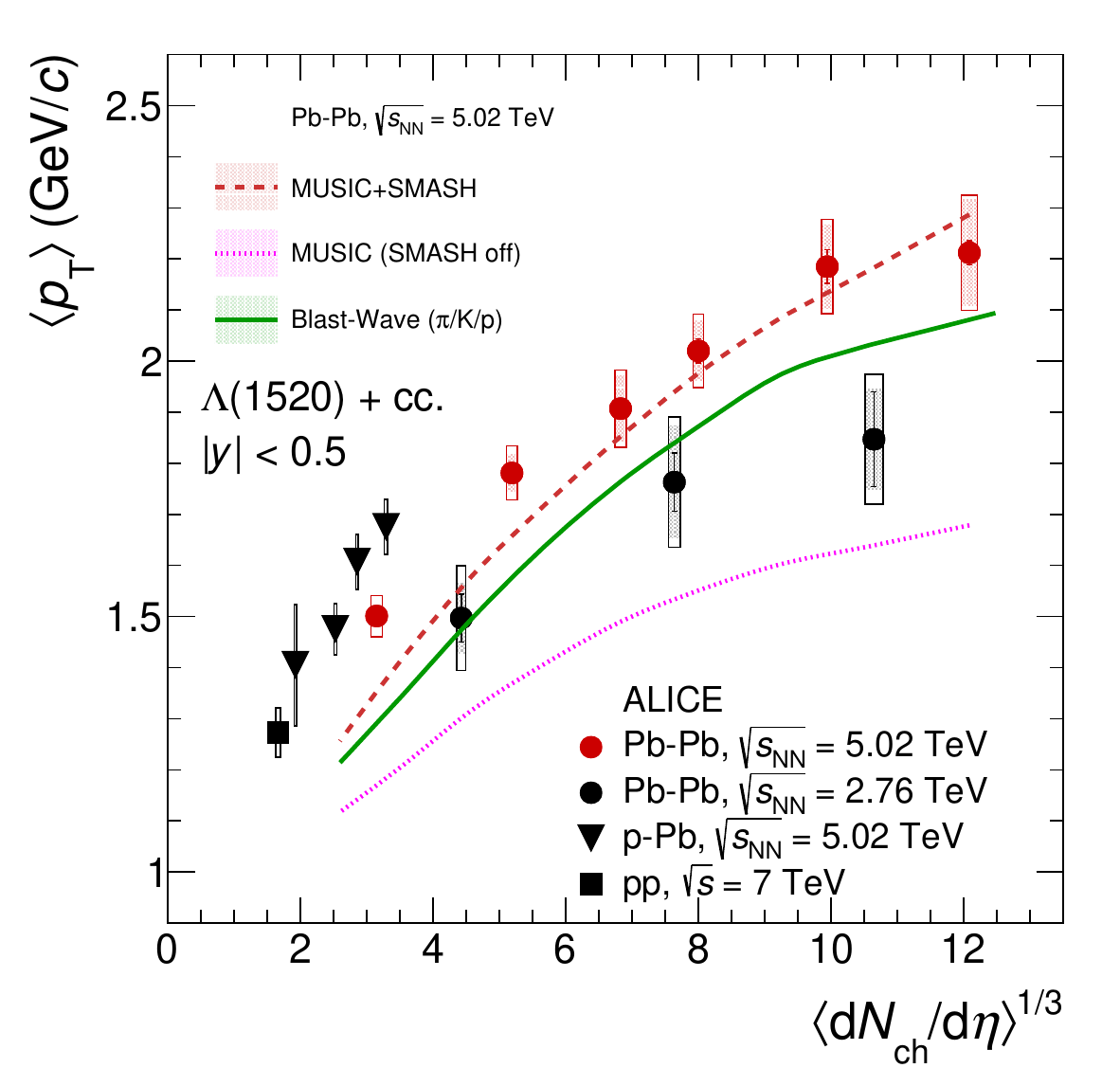}
\caption{The \average{\pt} of \Lambdastar as a function of \dnchdetacubic (red solid markers), compared with earlier data~\cite{ALICE:2018ewo,ALICE:2019smg}. Predictions from MUSIC with the SMASH afterburner~\cite{SMASH:2016zqf,Oliinychenko:2021enj}, and MUSIC without the afterburner are shown as red dashed and magenta dotted lines, respectively, for \PbPb collisions at \sqrtsNN = 5.02 TeV. The Blast-Wave predictions are shown as green solid lines. Statistical and systematic uncertainties are indicated by bars and boxes, respectively. Shaded boxes indicate uncorrelated systematic uncertainties.
}
\label{fig:meanpt}
\end{figure}

The \pt distributions of \Lambdastar, shown in Fig.~\ref{fig:spectra}, exhibit the characteristic hardening from peripheral to central \PbPb collisions observed for all light-flavor hadrons~\cite{ALICE:2019hno,ALICE:2019xyr}, a feature attributed to increasing radial flow and hard processes.
To test whether the \Lambdastar is emitted from the same expanding source as pions, kaons, and protons, the data are compared with Blast-Wave model predictions~\cite{Schnedermann:1993ws}. 
The model parameters, namely the average transverse flow velocity $\langle\beta_{\rm T}\rangle$, the kinetic freeze-out temperature $T_{\rm kin}$, and the exponent $n$ of the velocity profile, are fixed to the values obtained from simultaneous fits to the $\pi$/K/p spectra~\cite{ALICE:2019hno}, while the normalization is left as a free parameter. 
The common kinetic freeze-out temperature is an effective description within the Blast-Wave framework and does not imply a simultaneous physical freeze-out of all hadron species. The Blast-Wave predictions (solid lines in Fig.~\ref{fig:spectra}) describe the \Lambdastar spectra at intermediate \pt, supporting the scenario that the \Lambdastar takes part in the same collective expansion as $\pi$/K/p, with a transverse flow velocity increasing from peripheral to central collisions~\cite{ALICE:2019hno}. However, the data are systematically harder than the predictions: they lie below the model at low \pt and above it at high \pt, particularly in peripheral collisions, where hard-scattering contributions not included in the model become relevant. Since the Blast-Wave model does not include the hadronic rescattering that affects short-lived resonances after chemical freeze-out, the low-\pt deficit is consistent with the loss of reconstructable \Lambdastar due to rescattering of their decay products.

Figure~\ref{fig:spectra} compares the data with predictions from the MUSIC hydrodynamic model~\cite{SMASH:2016zqf,Oliinychenko:2021enj}, both with and without the SMASH hadronic afterburner, which simulates the hadronic interactions occurring after the hydrodynamic expansion stage. The calculations are performed for particles only and scaled by a factor of two assuming particle--antiparticle symmetry. When the hadronic phase is simulated with SMASH, the model describes the measured spectra in all centrality classes up to 4.0~\GeVc and reproduces the observed centrality-dependent hardening. Without the SMASH stage (available for 0--70\% centrality and \pt~$<$~4~\GeVc), the model fails to describe the data, with the largest deviations at low \pt in central collisions, decreasing towards peripheral collisions. Since both calculations share the same hydrodynamic evolution, this difference directly isolates the effect of the hadronic phase, being strongest in central collisions at low \pt, where the longer-lived and denser hadronic phase enhances the rescattering of the \Lambdastar decay products.

The \pt-integrated yield, \dndy, and the average transverse momentum, \average{\pt}, are obtained by integrating the measured \Lambdastar spectra and extrapolating to unmeasured \pt regions using separate Blast-Wave fits to each \pt-distribution. In these fits, the normalization and shape parameters are free, unlike in the comparison above, where the shape parameters are fixed to those extracted from $\pi$/K/p spectra. The fits describe the \pt dependence of the data within the measured range.
The extrapolation fraction ranges from 1.3\% in central to 10.4\% in peripheral collisions.
Systematic uncertainties of the extrapolated yields are evaluated by fitting the spectral shapes with Fermi-Dirac, Maxwell-Boltzmann, \mt-exponential, Levy-Tsallis functions, and Blast-Wave functions with parameters fixed to the $\pi$/K/p fit~\cite{ALICE:2019hno}, yielding 5.8\%, 4.6\%, 3.4\%, 3.3\%, 2.6\%, and 5.2\%, respectively, for central to peripheral centralities.
The total relative systematic uncertainties on \dndy are 15.1\%, 13.0\%, 10.2\%, 10.9\%, 9.5\%, and 8.7\%, including a significant common 7.5\% across all centrality classes. Those on \average{\pt} are 8.7\%, 7.1\%, 5.6\%, 5.8\%, 4.9\%, and 3.9\%, including a common contribution of 4\%.

\begin{figure}
\centering
\includegraphics[width=0.5\textwidth]{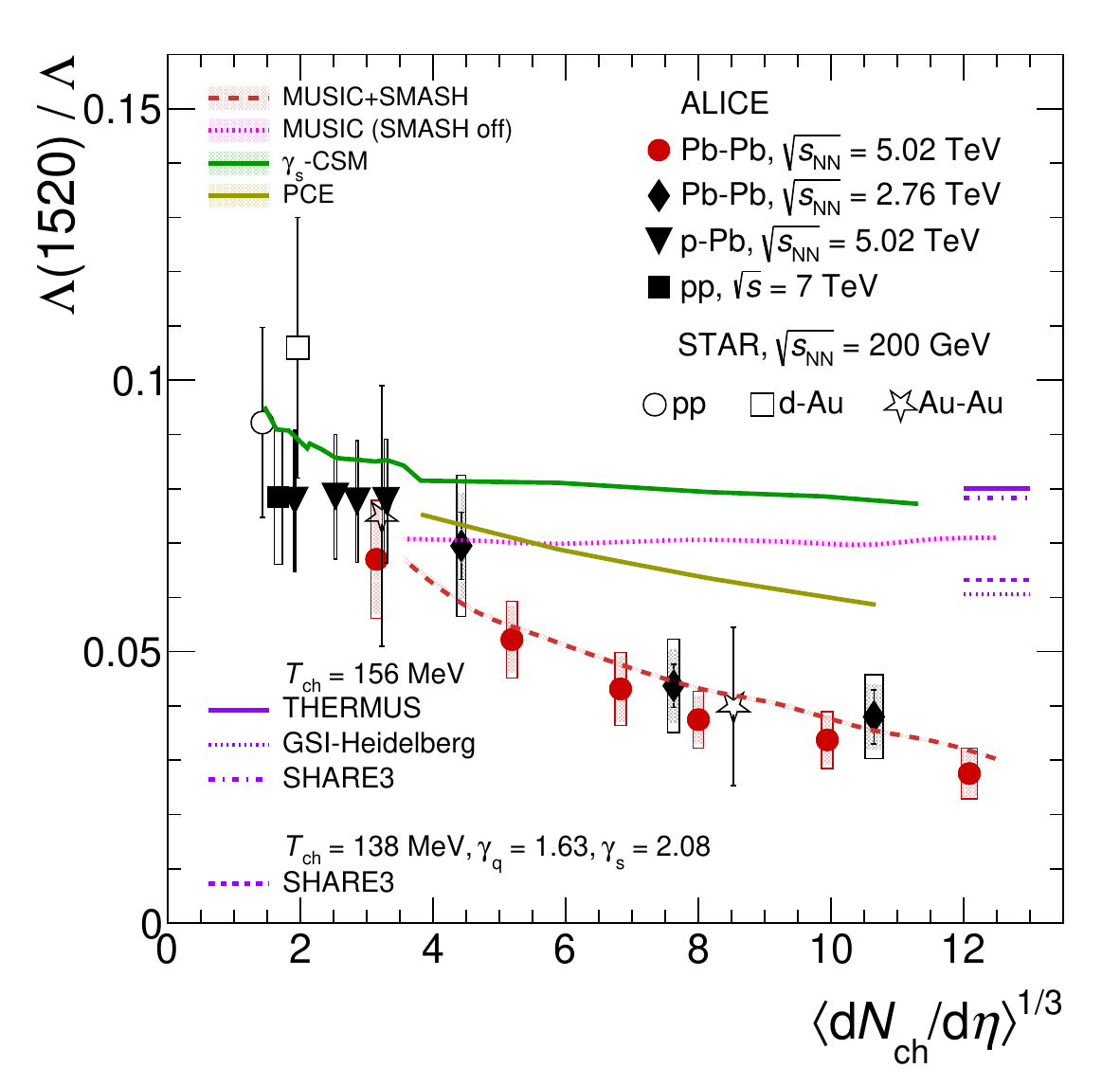}
\caption{The \pt-integrated \Lambdastar/\pLambda ratio as a function of \dnchdetacubic (red solid markers). Predictions from MUSIC with and without the SMASH afterburner are shown as red dashed and magenta dotted lines, respectively. CSM and PCE thermal model predictions are shown as green and yellow lines. The predictions from several SHMs are also included. Statistical and systematic uncertainties are indicated by bars and boxes, respectively. Shaded boxes indicate uncorrelated systematic uncertainties.
}
\label{fig:ratio}
\end{figure}

Figures~\ref{fig:meanpt} and~\ref{fig:ratio} show, respectively, the \average{\pt} of \Lambdastar and the \Lambdastar/\pLambda yield ratios in \PbPb collisions at \sqrtsNN = 5.02 TeV as a function of the cubic root of the charged-particle multiplicity density at midrapidity~\cite{Aamodt:2010cz}, \dnchdetacubic. 
Results are compared with pp~\cite{ALICE:2019smg}, p--Pb~\cite{ALICE:2019smg} and lower energy \PbPb~\cite{ALICE:2018ewo} data.

The \average{\pt} of \Lambdastar rises significantly with the charged-particle multiplicity, i.e., with collision centrality, reaching a value in central collisions that is about 47.4\% higher than in peripheral collisions.
Moreover, the \average{\pt} of \Lambdastar is significantly higher in \PbPb collisions at \sqrtsNN = 5.02 TeV than at 2.76 TeV, consistent with the stronger radial flow observed at the higher energy, also reflected in the $\pi$/K/p measurements~\cite{ALICE:2019hno}.
The Blast-Wave model~\cite{Schnedermann:1993ws,Abelev:2013vea} (green solid line in Fig.~\ref{fig:meanpt}) reproduces the rising trend of \average{\pt} with centrality but underestimates the data by 6–20\% from central to peripheral collisions. The centrality dependence of \average{\pt} is not fully captured by MUSIC+SMASH (red dashed line), which agrees with data in central and semi-central collisions but underestimates the 50--70\% and 70--90\% centrality classes by 6.7–9.2\%. Strikingly, without the SMASH afterburner (magenta dotted line), \average{\pt} values lie significantly below the data across all centralities, highlighting the importance of including hadronic interactions in modelling the evolution of heavy-ion collisions.

The ratio of the \pt-integrated yield of \Lambdastar to the longer-lived ground state \pLambda~\cite{ALICE:2025cqy} provides a centrality-dependent probe of particle production, while cancelling effects due to valence-quark composition, such as the strangeness enhancement~\cite{ALICE:2016fzo,ALICE:2019avo}, and highlighting the role of the short resonance lifetime. For the yield ratio, both the resonance and ground-state yields are the sum of particles and antiparticles; the ground-state yields are taken from Ref.~\cite{ALICE:2025cqy}.

As shown in Fig.~\ref{fig:ratio}, the \Lambdastar/\pLambda ratio decreases with increasing charged-particle multiplicity. The suppression is quantified through the double ratio of yield ratios between central and peripheral collisions, 
\begin{equation}
R^{\Lambdastar/\pLambda}_{\rm cp} = \frac{\left[\Lambdastar/\Lambda\right]_{\rm \, 0-10\% \,\,\,\,\, }}{\left[\Lambdastar/\Lambda\right]_{\rm \, 70-90\%}} = 0.411 \pm 0.015~({\rm stat}) \pm 0.086~({\rm syst}), \nonumber
\end{equation}

where the uncertainties that are correlated across centrality (see Table~\ref{tab:sys_summary}) cancel. A suppression of $(59 \pm 1.5^{stat} \pm 8.6^{syst})\%$ is observed for the \Lambdastar/\pLambda ratio in \PbPb collisions at \sqrtsNN = 5.02 TeV. 
With a significance of 6.8$\sigma$, this represents the first observation above 5$\sigma$ of resonance suppression in central relative to peripheral collisions within the same collision system. For the \Kstarpm, the suppression in central Pb–Pb collisions reaches 9.3$\sigma$ relative to pp collisions~\cite{ALICE:2023ifn}, while the corresponding central-to-peripheral suppression remains below 5$\sigma$.
Previous ALICE results at \sqrtsNN = 2.76 TeV (black round markers in Fig.~\ref{fig:ratio}) reported a $(45 \pm 8^{stat} \pm 12^{syst})\% $ suppression between central (0--20\%) and peripheral (50--80\%) collisions, with 3.1$\sigma$ significance. 
The STAR measurement in \AuAu, \dAu, and pp collisions at \sqrtsNN~=~200 GeV~\cite{Adams:2006yu,Abelev:2008yz}, also reported in Fig.~\ref{fig:ratio}, follow a similar trend but shows only 1.8$\sigma$ significance for suppression in \AuAu relative to pp. The presented result at \mbox{\sqrtsNN = 5.02 TeV} agrees with the 2.76 TeV measurement within uncertainties, while offering finer centrality intervals and substantially improved precision.

The multiplicity-dependent suppression trend of the \Lambdastar/\pLambda ratio is reproduced by MUSIC+SMASH predictions (red dashed line in Fig.~\ref{fig:ratio}), in good agreement with the data within uncertainties. The predicted double ratio $R^{\Lambdastar/\pLambda}_{\rm cp}$ is about 25\% larger than the measured value, but it remains consistent with data within uncertainties. Notably, without the SMASH afterburner (magenta dotted line), the model yields \Lambdastar/\pLambda values consistent with peripheral \PbPb, p–Pb, and pp collisions, showing no centrality dependence.
Comparing MUSIC with and without the SMASH afterburner demonstrates that incorporating interactions in the late hadronic phase is essential for a realistic description of heavy-ion collisions, enabling the model to reproduce the \Lambdastar yield, \average{\pt}, and the observed centrality-dependent suppression of the \Lambdastar/\pLambda ratio.

The latter is compared with predictions from statistical hadronisation models (SHMs). The canonical statistical hadronization model (CSM, green solid line in Fig.~\ref{fig:ratio})~\cite{Vovchenko:2019kes}, based on a fixed chemical freeze-out temperature $T_{\rm ch}$ = 155 MeV and a correlation volume of three units of rapidity ($V_{c}$ = 3d$V$/d$y$), reproduces the hadron-to-pion ratios across the ALICE multiplicity ranges, except for the $\phi/\pi$ ratio. To account for this, incomplete strangeness saturation is introduced via a factor $\gamma_{\mathrm{s}}$, and its predictions for resonance-to-stable particle yield ratios are in good agreement with the ALICE  multiplicity-dependent measurements (Fig. 6 of Ref. ~\cite{Vovchenko:2019kes}), as the effects of baryon-number conservation and incomplete strangeness equilibration largely cancel. The $\gamma_{\mathrm{S}}$-CSM model (green line in Fig.~\ref{fig:ratio}) predicts systematically higher \Lambdastar/\pLambda ratios, showing a weak multiplicity dependence. Thermal-model calculations with partial chemical equilibrium (PCE, yellow line in Fig~\ref{fig:ratio})~\cite{Motornenko:2019jha} capture the suppression trend qualitatively but overestimate the absolute value. Grand-canonical statistical hadronisation models (GSI-Heidelberg~\cite{Andronic:2005yp,Andronic:2017pug}), THERMUS~\cite{Wheaton:2004vg} and SHARE3~\cite{Petran:2013dva} constrained by stable-hadron fits~\cite{Floris:2014pta}, as well as the non-equilibrium SHARE3 configuration with free $\gamma_{s}$ (strange) and $\gamma_{q}$  parameters~\cite{Petran:2013lja} are compatible with the \Lambdastar/\pLambda ratio measured in most peripheral collisions within uncertainties. The \Lambdastar/\pLambda ratio in most central collisions is about 55\%, 66\%, 65\%, and 56\% lower than calculations from GSI-Heidelberg, THERMUS, SHARE3 (both equilibrium and non-equilibrium configuration) models, respectively. As SHMs describe generally well the particle ratios at $T_{\rm ch}$ = 156 MeV, this comparison shows that additional interactions occur between chemical and kinetic freeze-out, such as hadronic rescattering, effectively reducing the yield of reconstructed resonances in central collisions. The observed centrality-dependence suggests that the magnitude of this effect is sensitive to the hadronic phase lifetime, which increases with centrality~\cite{ALICE:2011dyt}. Overall, the observed suppression and model comparison strongly support the existence of a hadronic phase that is sufficiently long-lived for rescattering to significantly modify the final-state \Lambdastar yield.
 
In conclusion, the presented measurement of \Lambdastar production in \PbPb collisions at \sqrtsNN = 5.02 TeV provides unprecedented granularity in both centrality and \pt and constitutes the most precise and significant benchmark to date in heavy-ion collision studies.
The spectral shapes are consistent with a hydrodynamic evolution, as observed for $\pi$, K, and p, indicating that the \Lambdastar takes part in the collective radial expansion. 
MUSIC coupled to the SMASH hadronic afterburner describes the measured spectra and \average{\pt} in central and semi-central classes, while discrepancies remain in the most peripheral ones. The calculation without hadronic transport overestimates the low-\pt yields and underestimates \average{\pt}. The \pt-integrated \Lambdastar/\pLambda ratio is suppressed by approximately 59\% in central (0--10\%) relative to peripheral (70--90\%) collisions with a significance of 6.8$\sigma$, which represents the first observation above 5$\sigma$ of resonance suppression within the same collision system.
Its centrality dependence is reproduced by MUSIC+SMASH within experimental uncertainties, whereas the calculation without hadronic transport predicts little centrality dependence. These comparisons support a substantial role for late-stage hadronic interactions in determining the reconstructible \Lambdastar yield.
The precision and granularity of the presented \Lambdastar/\pLambda measurement provide a stringent test of hadronic transport calculations through the combined study of yield suppression and spectral evolution. Together with the model comparisons, these results offer new constraints on late-stage hadronic dynamics and provide further evidence for the existence of a hadronic phase in heavy-ion collisions.


\newenvironment{acknowledgement}{\relax}{\relax}
\begin{acknowledgement}
\section*{Acknowledgements}

The ALICE Collaboration would like to thank all its engineers and technicians for their invaluable contributions to the construction of the experiment and the CERN accelerator teams for the outstanding performance of the LHC complex.
The ALICE Collaboration gratefully acknowledges the resources and support provided by all Grid centres and the Worldwide LHC Computing Grid (WLCG) collaboration.
The ALICE Collaboration acknowledges the following funding agencies for their support in building and running the ALICE detector:
A. I. Alikhanyan National Science Laboratory (Yerevan Physics Institute) Foundation (ANSL), State Committee of Science and World Federation of Scientists (WFS), Armenia;
Austrian Academy of Sciences, Austrian Science Fund (FWF): [M 2467-N36] and Nationalstiftung f\"{u}r Forschung, Technologie und Entwicklung, Austria;
Ministry of Communications and High Technologies, National Nuclear Research Center, Azerbaijan;
Rede Nacional de Física de Altas Energias (Renafae), Financiadora de Estudos e Projetos (Finep), Funda\c{c}\~{a}o de Amparo \`{a} Pesquisa do Estado de S\~{a}o Paulo (FAPESP) and The Sao Paulo Research Foundation  (FAPESP), Brazil;
Bulgarian Ministry of Education and Science, within the National Roadmap for Research Infrastructures 2020-2027 (object CERN), Bulgaria;
Ministry of Education of China (MOEC) , Ministry of Science \& Technology of China (MSTC) and National Natural Science Foundation of China (NSFC), China;
Ministry of Science and Education and Croatian Science Foundation, Croatia;
Centro de Aplicaciones Tecnol\'{o}gicas y Desarrollo Nuclear (CEADEN), Cubaenerg\'{\i}a, Cuba;
Ministry of Education, Youth and Sports of the Czech Republic, Czech Republic;
The Danish Council for Independent Research | Natural Sciences, the VILLUM FONDEN and Danish National Research Foundation (DNRF), Denmark;
Helsinki Institute of Physics (HIP), Finland;
Commissariat \`{a} l'Energie Atomique (CEA) and Institut National de Physique Nucl\'{e}aire et de Physique des Particules (IN2P3) and Centre National de la Recherche Scientifique (CNRS), France;
Bundesministerium f\"{u}r Forschung, Technologie und Raumfahrt (BMFTR) and GSI Helmholtzzentrum f\"{u}r Schwerionenforschung GmbH, Germany;
National Research, Development and Innovation Office, Hungary;
Department of Atomic Energy Government of India (DAE), Department of Science and Technology, Government of India (DST), University Grants Commission, Government of India (UGC) and Council of Scientific and Industrial Research (CSIR), India;
National Research and Innovation Agency - BRIN, Indonesia;
Istituto Nazionale di Fisica Nucleare (INFN), Italy;
Japanese Ministry of Education, Culture, Sports, Science and Technology (MEXT) and Japan Society for the Promotion of Science (JSPS) KAKENHI, Japan;
Consejo Nacional de Ciencia (CONACYT) y Tecnolog\'{i}a, through Fondo de Cooperaci\'{o}n Internacional en Ciencia y Tecnolog\'{i}a (FONCICYT) and Direcci\'{o}n General de Asuntos del Personal Academico (DGAPA), Mexico;
Nederlandse Organisatie voor Wetenschappelijk Onderzoek (NWO), Netherlands;
The Research Council of Norway, Norway;
Pontificia Universidad Cat\'{o}lica del Per\'{u}, Peru;
Ministry of Science and Higher Education, National Science Centre and WUT ID-UB, Poland;
Korea Institute of Science and Technology Information and National Research Foundation of Korea (NRF), Republic of Korea;
Ministry of Education and Scientific Research, Institute of Atomic Physics, Ministry of Research and Innovation and Institute of Atomic Physics and Universitatea Nationala de Stiinta si Tehnologie Politehnica Bucuresti, Romania;
Ministerstvo skolstva, vyskumu, vyvoja a mladeze SR, Slovakia;
National Research Foundation of South Africa, South Africa;
Swedish Research Council (VR) and Knut \& Alice Wallenberg Foundation (KAW), Sweden;
European Organization for Nuclear Research, Switzerland;
Suranaree University of Technology (SUT), National Science and Technology Development Agency (NSTDA) and National Science, Research and Innovation Fund (NSRF via PMU-B B05F650021), Thailand;
Turkish Energy, Nuclear and Mineral Research Agency (TENMAK), Turkey;
National Academy of  Sciences of Ukraine, Ukraine;
Science and Technology Facilities Council (STFC), United Kingdom;
National Science Foundation of the United States of America (NSF) and United States Department of Energy, Office of Nuclear Physics (DOE NP), United States of America.
In addition, individual groups or members have received support from:
Czech Science Foundation (grant no. 23-07499S), Czech Republic;
FORTE project, reg.\ no.\ CZ.02.01.01/00/22\_008/0004632, Czech Republic, co-funded by the European Union, Czech Republic;
European Research Council (grant no. 101220549), European Union;
Deutsche Forschungs Gemeinschaft (DFG, German Research Foundation) ``Neutrinos and Dark Matter in Astro- and Particle Physics'' (grant no. SFB 1258), Germany;
CONVECS project, CUP C97H23001700002 FESR 2021-2027 program, Italy.

\end{acknowledgement}

\bibliographystyle{utphys}   
\bibliography{bibliography}

\newpage
\appendix

%
%

\section{The ALICE Collaboration}
\label{app:collab}
\begin{flushleft} 
\small

D.A.H.~Abdallah\,\orcidlink{0000-0003-4768-2718}\,$^{\rm 133}$, 
M.M.~Abedin$^{\rm 4}$, 
I.J.~Abualrob\,\orcidlink{0009-0005-3519-5631}\,$^{\rm 110}$, 
S.~Acharya\,\orcidlink{0000-0002-9213-5329}\,$^{\rm 49}$, 
A.B.~Adiguzel$^{\rm 90}$, 
K.~Agarwal\,\orcidlink{0000-0001-5781-3393}\,$^{\rm II,}$$^{\rm 23}$, 
G.~Aglieri Rinella\,\orcidlink{0000-0002-9611-3696}\,$^{\rm 32}$, 
L.~Aglietta\,\orcidlink{0009-0003-0763-6802}\,$^{\rm 24}$, 
N.~Agrawal\,\orcidlink{0000-0003-0348-9836}\,$^{\rm 25}$, 
Z.~Ahammed\,\orcidlink{0000-0001-5241-7412}\,$^{\rm 131}$
% ...

\section*{Affiliation Notes}

$^{\rm I}$ Deceased\\
$^{\rm II}$ Also at: INFN Trieste, Trieste, Italy\\
$^{\rm III}$ Also at: Fondazione Bruno Kessler (FBK), Trento, Italy\\
$^{\rm IV}$ Also at: Czech Technical University in Prague, Prague, Czech Republic\\
$^{\rm V}$ Also at: Dipartimento DET del Politecnico di Torino, Turin, Italy\\
$^{\rm VI}$ Also at: University College of Dublin, Dublin, Ireland\\
$^{\rm VII}$ Also at: Department of Applied Physics, Aligarh Muslim University, Aligarh, India\\
$^{\rm VIII}$ Also at: Institute of Theoretical Physics, University of Wroclaw, Wroclaw, Poland\\
$^{\rm IX}$ Also at: Facultad de Ciencias, Universidad Nacional Aut\'{o}noma de M\'{e}xico, Mexico City, Mexico\\
$^{\rm X}$ Also at: Lovely Professional University, Punjab, India
\\

\section*{Collaboration Institutes}

$^{1}$ A.I. Alikhanyan National Science Laboratory (Yerevan Physics Institute) Foundation, Yerevan, Armenia\\
$^{2}$ AGH University of Krakow, Cracow, Poland\\
$^{3}$ Bogolyubov Institute for Theoretical Physics, National Academy of Sciences of Ukraine, Kyiv, Ukraine\\
$^{4}$ Bose Institute, Department of Physics  and Centre for Astroparticle Physics and Space Science (CAPSS), Kolkata, India\\
$^{5}$ California Polytechnic State University, San Luis Obispo, California, United States\\
$^{6}$ Central China Normal University, Wuhan, China\\
$^{7}$ Centro de Aplicaciones Tecnol\'{o}gicas y Desarrollo Nuclear (CEADEN), Havana, Cuba\\
$^{8}$ Centro de Investigaci\'{o}n y de Estudios Avanzados (CINVESTAV), Mexico City and M\'{e}rida, Mexico\\
$^{9}$ Chicago State University, Chicago, Illinois, United States\\
$^{10}$ China Nuclear Data Center, China Institute of Atomic Energy, Beijing, China\\
$^{11}$ China University of Geosciences, Wuhan, China\\
$^{12}$ Chungbuk National University, Cheongju, Republic of Korea\\
$^{13}$ Comenius University Bratislava, Faculty of Mathematics, Physics and Informatics, Bratislava, Slovak Republic\\
$^{14}$ Creighton University, Omaha, Nebraska, United States\\
$^{15}$ Department of Physics, Aligarh Muslim University, Aligarh, India\\
$^{16}$ Department of Physics, Pusan National University, Pusan, Republic of Korea\\
$^{17}$ Department of Physics, Sejong University, Seoul, Republic of Korea\\
$^{18}$ Department of Physics, University of California, Berkeley, California, United States\\
$^{19}$ Department of Physics, University of Oslo, Oslo, Norway\\
$^{20}$ Department of Physics and Technology, University of Bergen, Bergen, Norway\\
$^{21}$ Dipartimento di Fisica, Universit\`{a} di Pavia, Pavia, Italy\\
$^{22}$ Dipartimento di Fisica dell'Universit\`{a} and Sezione INFN, Cagliari, Italy\\
$^{23}$ Dipartimento di Fisica dell'Universit\`{a} and Sezione INFN, Trieste, Italy\\
$^{24}$ Dipartimento di Fisica dell'Universit\`{a} and Sezione INFN, Turin, Italy\\
$^{25}$ Dipartimento di Fisica e Astronomia dell'Universit\`{a} and Sezione INFN, Bologna, Italy\\
$^{26}$ Dipartimento di Fisica e Astronomia dell'Universit\`{a} and Sezione INFN, Catania, Italy\\
$^{27}$ Dipartimento di Fisica e Astronomia dell'Universit\`{a} and Sezione INFN, Padova, Italy\\
$^{28}$ Dipartimento di Fisica `E.R.~Caianiello' dell'Universit\`{a} and Gruppo Collegato INFN, Salerno, Italy\\
$^{29}$ Dipartimento DISAT del Politecnico and Sezione INFN, Turin, Italy\\
$^{30}$ Dipartimento di Scienze MIFT, Universit\`{a} di Messina, Messina, Italy\\
$^{31}$ Dipartimento Interateneo di Fisica `M.~Merlin' and Sezione INFN, Bari, Italy\\
$^{32}$ European Organization for Nuclear Research (CERN), Geneva, Switzerland\\
$^{33}$ Faculty of Electrical Engineering, Mechanical Engineering and Naval Architecture, University of Split, Split, Croatia\\
$^{34}$ Faculty of Nuclear Sciences and Physical Engineering, Czech Technical University in Prague, Prague, Czech Republic\\
$^{35}$ Faculty of Physics, Sofia University, Sofia, Bulgaria\\
$^{36}$ Faculty of Science, P.J.~\v{S}af\'{a}rik University, Ko\v{s}ice, Slovak Republic\\
$^{37}$ Faculty of Technology, Environmental and Social Sciences, Bergen, Norway\\
$^{38}$ Frankfurt Institute for Advanced Studies, Johann Wolfgang Goethe-Universit\"{a}t Frankfurt, Frankfurt, Germany\\
$^{39}$ Fudan University, Shanghai, China\\
$^{40}$ Gauhati University, Department of Physics, Guwahati, India\\
$^{41}$ Helmholtz-Institut f\"{u}r Strahlen- und Kernphysik, Rheinische Friedrich-Wilhelms-Universit\"{a}t Bonn, Bonn, Germany\\
$^{42}$ Helsinki Institute of Physics (HIP), Helsinki, Finland\\
$^{43}$ High Energy Physics Group,  Universidad Aut\'{o}noma de Puebla, Puebla, Mexico\\
$^{44}$ Horia Hulubei National Institute of Physics and Nuclear Engineering, Bucharest, Romania\\
$^{45}$ HUN-REN Wigner Research Centre for Physics, Budapest, Hungary\\
$^{46}$ Indian Institute of Technology Bombay (IIT), Mumbai, India\\
$^{47}$ Indian Institute of Technology Indore, Indore, India\\
$^{48}$ INFN, Laboratori Nazionali di Frascati, Frascati, Italy\\
$^{49}$ INFN, Sezione di Bari, Bari, Italy\\
$^{50}$ INFN, Sezione di Bologna, Bologna, Italy\\
$^{51}$ INFN, Sezione di Cagliari, Cagliari, Italy\\
$^{52}$ INFN, Sezione di Catania, Catania, Italy\\
$^{53}$ INFN, Sezione di Padova, Padova, Italy\\
$^{54}$ INFN, Sezione di Pavia, Pavia, Italy\\
$^{55}$ INFN, Sezione di Torino, Turin, Italy\\
$^{56}$ INFN, Sezione di Trieste, Trieste, Italy\\
$^{57}$ Inha University, Incheon, Republic of Korea\\
$^{58}$ Institute for Gravitational and Subatomic Physics (GRASP), Utrecht University/Nikhef, Utrecht, Netherlands\\
$^{59}$ Institute of Experimental Physics, Slovak Academy of Sciences, Ko\v{s}ice, Slovak Republic\\
$^{60}$ Institute of Physics of the Czech Academy of Sciences, Prague, Czech Republic\\
$^{61}$ Institute of Space Science (ISS), Bucharest, Romania\\
$^{62}$ Institut f\"{u}r Kernphysik, Johann Wolfgang Goethe-Universit\"{a}t Frankfurt, Frankfurt, Germany\\
$^{63}$ Instituto de Ciencias Nucleares, Universidad Nacional Aut\'{o}noma de M\'{e}xico, Mexico City, Mexico\\
$^{64}$ Instituto de F\'{i}sica, Universidade Federal do Rio Grande do Sul (UFRGS), Porto Alegre, Brazil\\
$^{65}$ Instituto de F\'{\i}sica, Universidad Nacional Aut\'{o}noma de M\'{e}xico, Mexico City, Mexico\\
$^{66}$ iThemba LABS, National Research Foundation, Somerset West, South Africa\\
$^{67}$ Jeonbuk National University, Jeonju, Republic of Korea\\
$^{68}$ Korea Institute of Science and Technology Information, Daejeon, Republic of Korea\\
$^{69}$ Laboratoire de Physique Subatomique et de Cosmologie, Universit\'{e} Grenoble-Alpes, CNRS-IN2P3, Grenoble, France\\
$^{70}$ Lawrence Berkeley National Laboratory, Berkeley, California, United States\\
$^{71}$ Lund University Department of Physics, Division of Particle Physics, Lund, Sweden\\
$^{72}$ Marietta Blau Institute, Vienna, Austria\\
$^{73}$ Nagasaki Institute of Applied Science, Nagasaki, Japan\\
$^{74}$ Nara Women{'}s University (NWU), Nara, Japan\\
$^{75}$ National Centre for Nuclear Research, Warsaw, Poland\\
$^{76}$ National Institute of Science Education and Research, Homi Bhabha National Institute, Jatni, India\\
$^{77}$ National Nuclear Research Center, Baku, Azerbaijan\\
$^{78}$ National Research and Innovation Agency - BRIN, Jakarta, Indonesia\\
$^{79}$ Niels Bohr Institute, University of Copenhagen, Copenhagen, Denmark\\
$^{80}$ Nikhef, National institute for subatomic physics, Amsterdam, Netherlands\\
$^{81}$ Nuclear Physics Group, STFC Daresbury Laboratory, Daresbury, United Kingdom\\
$^{82}$ Nuclear Physics Institute of the Czech Academy of Sciences, Husinec-\v{R}e\v{z}, Czech Republic\\
$^{83}$ Oak Ridge National Laboratory, Oak Ridge, Tennessee, United States\\
$^{84}$ Ohio State University, Columbus, Ohio, United States\\
$^{85}$ Physics Department, Panjab University, Chandigarh, India\\
$^{86}$ Physics Department, University of Jammu, Jammu, India\\
$^{87}$ Physics Program and International Institute for Sustainability with Knotted Chiral Meta Matter (WPI-SKCM$^{2}$), Hiroshima University, Hiroshima, Japan\\
$^{88}$ Physikalisches Institut, Eberhard-Karls-Universit\"{a}t T\"{u}bingen, T\"{u}bingen, Germany\\
$^{89}$ Physikalisches Institut, Ruprecht-Karls-Universit\"{a}t Heidelberg, Heidelberg, Germany\\
$^{90}$ Physik Department, Technische Universit\"{a}t M\"{u}nchen, Munich, Germany\\
$^{91}$ Politecnico di Bari and Sezione INFN, Bari, Italy\\
$^{92}$ Research Division and ExtreMe Matter Institute EMMI, GSI Helmholtzzentrum f\"ur Schwerionenforschung GmbH, Darmstadt, Germany\\
$^{93}$ Saga University, Saga, Japan\\
$^{94}$ Saha Institute of Nuclear Physics, Homi Bhabha National Institute, Kolkata, India\\
$^{95}$ School of Physics and Astronomy, University of Birmingham, Birmingham, United Kingdom\\
$^{96}$ Secci\'{o}n F\'{\i}sica, Departamento de Ciencias, Pontificia Universidad Cat\'{o}lica del Per\'{u}, Lima, Peru\\
$^{97}$ SUBATECH, IMT Atlantique, Nantes Universit\'{e}, CNRS-IN2P3, Nantes, France\\
$^{98}$ Sungkyunkwan University, Suwon City, Republic of Korea\\
$^{99}$ Suranaree University of Technology, Nakhon Ratchasima, Thailand\\
$^{100}$ Technical University of Ko\v{s}ice, Ko\v{s}ice, Slovak Republic\\
$^{101}$ The Henryk Niewodniczanski Institute of Nuclear Physics, Polish Academy of Sciences, Cracow, Poland\\
$^{102}$ The University of Texas at Austin, Austin, Texas, United States\\
$^{103}$ Universidad Aut\'{o}noma de Sinaloa, Culiac\'{a}n, Mexico\\
$^{104}$ Universidade de S\~{a}o Paulo (USP), S\~{a}o Paulo, Brazil\\
$^{105}$ Universidade Estadual de Campinas (UNICAMP), Campinas, Brazil\\
$^{106}$ Universidade Federal do ABC, Santo Andre, Brazil\\
$^{107}$ Universitatea Nationala de Stiinta si Tehnologie Politehnica Bucuresti, Bucharest, Romania\\
$^{108}$ University of Cape Town, Cape Town, South Africa\\
$^{109}$ University of Derby, Derby, United Kingdom\\
$^{110}$ University of Houston, Houston, Texas, United States\\
$^{111}$ University of Jyv\"{a}skyl\"{a}, Jyv\"{a}skyl\"{a}, Finland\\
$^{112}$ University of Kansas, Lawrence, Kansas, United States\\
$^{113}$ University of Liverpool, Liverpool, United Kingdom\\
$^{114}$ University of Science and Technology of China, Hefei, China\\
$^{115}$ University of Silesia in Katowice, Katowice, Poland\\
$^{116}$ University of South-Eastern Norway, Kongsberg, Norway\\
$^{117}$ University of Tennessee, Knoxville, Tennessee, United States\\
$^{118}$ University of the Witwatersrand, Johannesburg, South Africa\\
$^{119}$ University of Tokyo, Tokyo, Japan\\
$^{120}$ University of Tsukuba, Tsukuba, Japan\\
$^{121}$ University of Zagreb Faculty of Science, Department of Physics, Zagreb, Croatia\\
$^{122}$ Universit\"{a}t M\"{u}nster, Institut f\"{u}r Kernphysik, M\"{u}nster, Germany\\
$^{123}$ Universit\'{e} Clermont Auvergne, CNRS/IN2P3, LPC, Clermont-Ferrand, France\\
$^{124}$ Universit\'{e} de Lyon, CNRS/IN2P3, Institut de Physique des 2 Infinis de Lyon, Lyon, France\\
$^{125}$ Universit\'{e} de Strasbourg, CNRS, IPHC UMR 7178, F-67000 Strasbourg, France, Strasbourg, France\\
$^{126}$ Universit\'{e} Paris-Saclay, Centre d'Etudes de Saclay (CEA), IRFU, D\'{e}partment de Physique Nucl\'{e}aire (DPhN), Saclay, France\\
$^{127}$ Universit\'{e}  Paris-Saclay, CNRS/IN2P3, IJCLab, Orsay, France\\
$^{128}$ Universit\`{a} degli Studi di Foggia, Foggia, Italy\\
$^{129}$ Universit\`{a} del Piemonte Orientale, Vercelli, Italy\\
$^{130}$ Universit\`{a} di Brescia, Brescia, Italy\\
$^{131}$ Variable Energy Cyclotron Centre, Homi Bhabha National Institute, Kolkata, India\\
$^{132}$ Warsaw University of Technology, Warsaw, Poland\\
$^{133}$ Wayne State University, Detroit, Michigan, United States\\
$^{134}$ Yale University, New Haven, Connecticut, United States\\
$^{135}$ Yildiz Technical University, Istanbul, Turkey\\
$^{136}$ Yonsei University, Seoul, Republic of Korea\\
$^{137}$ Affiliated with an institute formerly covered by a cooperation agreement with CERN\\
$^{138}$ Affiliated with an international laboratory covered by a cooperation agreement with CERN.\\

\end{flushleft} 
  
\end{document}